\documentclass{ws-ijmpe}

\newcommand{\be}{\begin{equation}}
\newcommand{\ee}{\end{equation}}
\newcommand{\beq}{\begin{eqnarray}}
\newcommand{\eeq}{\end{eqnarray}}
\newcommand{\ba}{\begin{array}}
\newcommand{\ea}{\end{array}}

\begin{document}

\markboth{M. Rafalski, W. Satu{\l}a, J. Dobaczewski}
{Isospin mixing of isospin-projected Slater determinants: formalism and
preliminary applications}

\catchline{}{}{}{}{}

\title{Isospin mixing of isospin-projected Slater determinants:  \\
   formalism and preliminary applications}

\author{M. Rafalski$^a$, W. Satu{\l}a$^a$ and J. Dobaczewski$^{a,b}$}
\address{
$^a$Institute of Theoretical Physics, University of Warsaw\\
         ul. Ho\.za 69,00-681 Warsaw, Poland \\
$^b$Department of Physics, P.O. Box 35 (YFL),
FI-40014 University of Jyv\"askyl\"a, Finland}

\maketitle

\begin{history}
\received{(received date)}
\revised{(revised date)}
\end{history}

\begin{abstract}
We report on the development of a new theoretical tool that allows for isospin
projection of Slater determinants and we present its
first applications. In particular, we determine
the isospin mixing in ground states of $N=Z$ nuclei
and discuss its dependence on the size of the harmonic-oscillator basis
used in the calculations.
We also discuss the unphysical contribution
to the isospin mixing caused by the spurious isospin-symmetry breaking
inherent to the mean-field approach. We show that these contributions may
be as large as 30\% of the value of the isospin-mixing parameter.
\end{abstract}

\section{Introduction}

\noindent
Self-consistent mean-field (MF) approach is practically the only formalism allowing
for large-scale no-core computations in heavy open-shell nuclei with many
valence particles. Inherent to the MF approach is the mechanism of spontaneous symmetry
breaking, which is essentially the only way to allow for incorporating a
significant part of many-body correlations into a single intrinsic
(symmetry-breaking) Slater determinant. However, unlike in the cases of
rotational or translational symmetry-breaking schemes, violation of the
isobaric symmetry has two distinctively different sources. The unwanted or
{\it unphysical\/} source pertains directly to the MF
approximation\cite{[Eng70],[Cau80],[Cau82],[Aue83]}.
It manifests itself very clearly in the
ground-state wave functions calculated by using isospin-invariant
interactions, like Skyrme or Gogny forces, with Coulomb force neglected.
Indeed, such calculations manifestly break the isospin symmetry in all but the
$N=Z$ systems, simply because the self-consistent proton and neutron wave
functions are then different. The second source of the isospin-symmetry
violation is of strictly {\it physical\/} nature and is caused mostly by the
Coulomb field and, to a much lesser extent, by strong-force
isospin-non-invariant components.

Hereby, we report on the development of a new theoretical tool that
allows for isospin projection (after variation) of Slater
determinants. It has been implemented within the Hartree-Fock code
HFODD\cite{[Dob00d],[Dob04fw]}. First, the Slater determinants are determined
in a standard way by minimizing the Skyrme functional plus the
Coulomb energy. Both direct and exchange terms of the Coulomb energy
are calculated exactly. We allow for arbitrary spatial deformations of these
intrinsic states. Second, the isospin-projected components are
determined, and third, they are mixed so as to rediagonalize the
total Skyrme-plus-Coulomb Hamiltonian.

Such a three-step procedure allows, respectively, for (i) taking into account
the competition between the nuclear and Coulomb interactions in building up
the single Slater determinant, which is becoming 'deformed' in the isospace,
(ii) restoring the isospin symmetry, and (iii)
letting the nuclear and Coulomb interactions pick the correct
mixtures of symmetry-restored eigenstates of the isospin.

In the present study, we briefly overview the main theoretical
building blocks of the formalism (Sect.~\ref{section-theory}) and
discuss preliminary applications. In particular, we give results for
the isospin-mixing parameters ($\alpha_C$) calculated in the ground
states of $N=Z$ nuclei (Sect.~\ref{section-N0}). The ultimate goal
will be to perform simultaneous isospin and angular
momentum\cite{[Zdu07],[Zdu07a]} projections and to systematically
calculate the isospin-symmetry breaking corrections
to the Fermi matrix element ($\delta_C$) for the set of
nuclei undergoing the superallowed $0^+  \rightarrow 0^+$ Fermi beta
decay\cite{[Har05],[Tow08]}.

\section{Theoretical formalism: isospin restoration and Coulomb
rediagonalization scheme} \label{section-theory}

The first step and the starting point of our approach is the determination
of the isospin-symmetry-broken
single-particle (s.p.) Slater determinant   $|\textrm{HF} \rangle$
calculated by using the Hartree-Fock (HF) theory including the isospin-invariant Skyrme
($\hat V^S$) and the isospin-symmetry-breaking Coulomb ($\hat V^C$)
interactions:
\be\label{ham}
     \hat H = \hat H^S  +  \hat V^C \quad \textrm{where} \quad
     \hat H^S =\hat T + \hat V^S.
\ee
The isospace-deformed state  $|\textrm{HF} \rangle$
admixes higher  isospin components $T\geq |T_z|$:
\be\label{mix}
|\textrm{HF} \rangle = \sum_{T\geq |T_z|}b_{T,T_z}|\eta; T,T_z\rangle ,
\ee
where  $T$ and $T_z$ are the total isospin and its third component,
respectively, $\eta$ labels all other quantum numbers
pertaining to the $|\textrm{HF} \rangle$ state, and the coefficients
$b_{T,T_z}$ are such that $\sum_{T\geq |T_z|} |b_{T,T_z}|^2 = 1$.

In the second step we create the good-isospin states $|\eta; T,T_z\rangle$
by projecting them out from the Slater determinant $|\textrm{HF} \rangle$:
\be
      |\eta ; T,T_z\rangle =  \frac{1}{b_{T,T_z}} \hat P^T_{T_z T_z}|\textrm{HF}
      \rangle .
\ee
In the following, we denote the mixing coefficients $b_{T,T_z}$ and average energies
$E^{\textrm{BR}}_{T,T_z}$:
\be
\label{before}
                     |b_{T,T_z}|^2 = \langle \textrm{HF} | \hat P^T_{T_z
                     T_z}|\textrm{HF} \rangle
                           \quad , \quad
 E^{\textrm{BR}}_{T,T_z} =
\langle\eta ; T,T_z| \hat{H}| \eta ; T,T_z\rangle,
\ee
as being obtained {\it before rediagonalization}.
In the above formulae, $\hat P^T_{T_z T_z}$
denotes the conventional\cite{[Rin80]} SO(3) projection
operator reduced to one dimension due to the $T_z$ quantum number
conservation, that is:
\be
\hat P^T_{T_zT_z} = \frac{2T+1}{2} \int_0^\pi d\beta \sin\beta
d^{T}_{T_zT_z}(\beta) \hat R(\beta),
\ee
where $\hat R(\beta )  = e^{-i \beta \hat T_y}$
denotes active-rotation operator by the Euler angle $\beta$ in
the isospace and $d^{T}_{T_zT_z} (\beta)$ is the Wigner
$d$-function\cite{[Var88]}.

In the third step we mix the projected states,
\be\label{mix2}
|\eta; n,T_z\rangle = \sum_{T\geq |T_z|}a^n_{T,T_z}|\eta; T,T_z\rangle ,
\ee
and determine the mixing coefficients $a^n_{T,T_z}$ by diagonalizing
Hamiltonian (\ref{ham}) in the space of projected states,
\be\label{mix3}
\sum_{T'\geq |T_z|}\langle\eta; T,T_z|\hat{H}|\eta; T',T_z\rangle
a^n_{T',T_z}
 = E^{\textrm{AR}}_{n,T_z}a^n_{T,T_z},
\ee
where $n$ enumerates the obtained eigenstates.
In the following, we denote the mixing coefficients $a^n_{T,T_z}$ and eigenenergies
$E^{\textrm{AR}}_{n,T_z}$ as being obtained {\it after rediagonalization}.
The lowest-energy solution, for $n=1$, corresponds to the isospin mixing in the
ground state.

The Skyrme Hamiltonian, $\hat H^S$, is an isoscalar operator; hence, it contributes
only to the diagonal matrix elements of the Hamiltonian (\ref{ham}),
$\langle \eta; T,T_z | \hat H^S
| \eta; T,T_z \rangle$, which can be obtained from:
\be
\label{kernel}
        \langle\textrm{HF}| \hat H^{S} \hat
        P^T_{T_zT_z}|\textrm{HF}\rangle
     =  \int_0^\pi d\beta
        \sin\beta\, d^{T}_{T_zT_z}(\beta)
        \langle\textrm{HF}| \hat H^{S} \hat R(\beta)|\textrm{HF}\rangle .
\ee
Similarly, calculation of the diagonal and non-diagonal
matrix elements of the Coulomb interaction,
$\langle \eta; T,T_z | \hat V^C | \eta; T',T_z \rangle$,
can be efficiently performed after decomposing $\hat V^{C}$
into the isoscalar, $\hat V^{C}_{00}$, isovector, $\hat V^{C}_{10}$, and
isotensor, $\hat V^{C}_{20}$, components,
and by making use of the SO(3) transformation rules for the
spherical tensors under rotations in the isospace\cite{[Var88]}.
In the particular case of one-dimensional projection we deal with in
this work, all matrix elements of axial spherical tensors
reduce to one-dimensional integrals over the Euler angle $\beta$:
\beq
 \langle \textrm{HF}|
 \hat{P}^{T}_{T_z T_z} \hat{V}^C_{\lambda 0} \hat{P}^{T^\prime}_{T_z T_z}
 |\textrm{HF} \rangle
      & = & C^{T T_z}_{ T^\prime T_z\; \lambda 0}
      \sum_{\mu' = -\lambda}^{\lambda}
      C^{T T_z}_{ T^\prime T^\prime_z \;  \lambda \mu' } \nonumber \\
      \frac{2T^\prime +1}{2} \int_0^\pi & d\beta & \;  \sin\beta\;
       d^{T^\prime}_{T^\prime_z, T_z} (\beta )\; \langle \textrm{HF}| \hat
       V^C_{\lambda \mu'} \hat R
      (\beta) |\textrm{HF} \rangle ,
\eeq
where $T^\prime_z = T_z-\mu'$ and
$C^{T T_z}_{ T^\prime T^\prime_z \;  \lambda \mu }$ denote standard
Clebsch-Gordan coefficients.
The Skyrme-Hamiltonian and Coulomb-interaction kernels,
$\langle\textrm{HF}| \hat H^{S} \hat R(\beta)|\textrm{HF}\rangle$ and
$\langle\textrm{HF}| \hat V^{C} \hat R(\beta)|\textrm{HF}\rangle$,
respectively, can be evaluated by using expressions for the standard
diagonal kernels\cite{[Per04]} ($\beta=0$) and replacing there the
isoscalar and isovector densities and currents with the so-called
transition densities and currents. Exact direct and exchange kernels
of the Coulomb interaction can be evaluated by using methods outlined
in Refs.\cite{[Gir83],[Dob96],[Dob05e]}.

\section{Numerical applications: the isospin mixing and
the isospin-projected energies in N=Z nuclei}\label{section-N0}

The isospin-mixing parameter, calculated before and after rediagonalization, is defined as
$\alpha_C = 1- |b_{|T_z|,T_z}|^2$ and $\alpha_C = 1- |a^{n=1}_{|T_z|,T_z}|^2$,
respectively. Its theoretical accuracy
depends on different factors, and in particular, on the size
of the spherical harmonic-oscillator (HO) basis used in the calculations.
A choice of the number
of the HO shells $N_0$ included in such calculations is always a result of
a trade-off between the accuracy and the CPU-time efficiency. In this respect, a
bottle-neck in our calculation scheme is the exact treatment of the exchange
Coulomb contribution, which makes calculations prohibitively
time consuming for $N_0 > 16 $.

\begin{figure}[t]
\begin{center}
\includegraphics[width=0.5\textwidth]{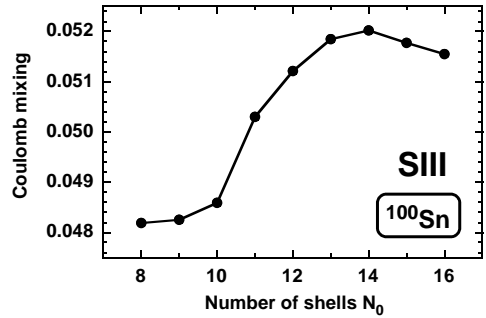}
\caption{The isospin-mixing parameter $\alpha_C$ calculated in $^{100}$Sn
(after rediagonalization)
as a function of the number of the HO shells $N_0$.
The results were obtained by using the SIII\protect\cite{[Bei75fw]} Skyrme
parameterization.
\label{mix-coul}}
\end{center}
\end{figure}

Dependence of the isospin-mixing parameter on $N_0$ is depicted in Fig.~\ref{mix-coul}.
The figure shows $\alpha_C$ in $^{100}$Sn, calculated after rediagonalization,
by using
the SIII Skyrme parameterization of Ref.\cite{[Bei75fw]}.
In the expanded scale of the figure, a significant variation
of the mixing parameter with $N_0$ is clearly seen.
Unfortunately, the
mixing parameter $\alpha_C$ does not stabilize at $N_0=16$.
Hence, by using the present method, $\alpha_C$ cannot be calculated
with the absolute precision greater than $\pm 0.002$,
or with the relative precision grater than $\pm 4$\%.
However, our studies show that the inaccuracy in evaluating $\alpha_C$
due to the basis cut-off appears to be much smaller than
the uncertainty related to the Skyrme force parameterization\cite{[Sat08a]}.

\begin{figure}[t]
\begin{center}
\includegraphics[width=0.7\textwidth ,clip=]{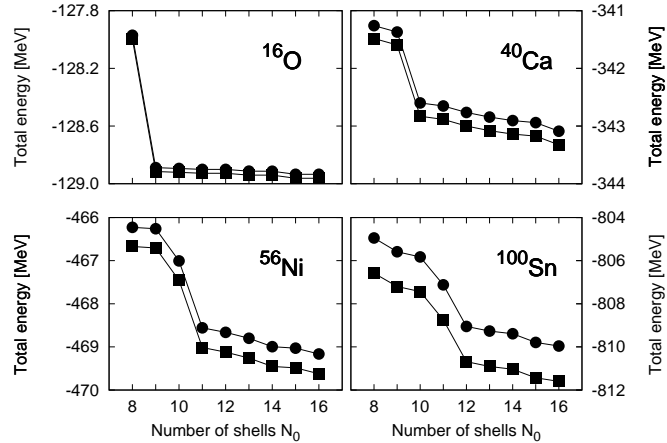}
\caption{Total binding energies in $^{16}$O, $^{40}$Ca, $^{56}$Ni, and
$^{100}$Sn as a functions of the number of the HO shells $N_0$.
Calculations were performed by using the SII Skyrme
force\protect\cite{[Bei75fw]}. Dots and squares label the
binding energies calculated before and after rediagonalization, respectively.
\label{SII}}
\end{center} \end{figure}

Fig.~\ref{SII} shows the total binding energies versus $N_0$, calculated
for doubly magic nuclei:
$^{16}$O, $^{40}$Ca, $^{56}$Ni, and $^{100}$Sn.
This set of calculations was performed by using the SII Skyrme force
parameterization\cite{[Bei75fw]}. The curves labeled by black squares
depict the projected energies $E^{\textrm{BR}}_{T,T_z}$
(\ref{before}), calculated before rediagonalization, and those marked
by triangles show the total binding energies $E^{\textrm{AR}}_{n=1,T_z}$
(\ref{mix3}), obtained after rediagonalization of the total
Hamiltonian in the isospin-projected basis. The figure shows that
({\it i}) the Coulomb rediagonalization effect increases with
increasing $Z$ as anticipated, and that ({\it ii}) the choice of $N_0
=12$ HO shells provides a reasonable estimate for the total binding
energy even for $^{100}$Sn. Hence, all calculations presented below
are done for $N_0 =12$.

\begin{figure}[t]
\begin{center}
\includegraphics[width=0.7\textwidth ,clip]{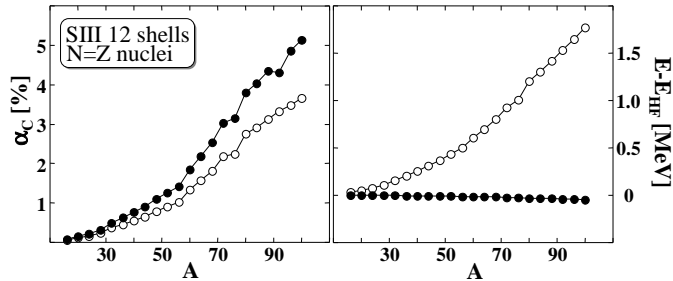}
\caption{The isospin mixing (left) and
the difference between the total binding energy and the HF energy
(right) in $N=Z$ nuclei, calculated for the SIII Skyrme force and $N_0=12$ HO shells.
The results shown by open and full dots
represent variants of the calculation before and after
rediagonalization, respectively.
\label{mixN=Z}}
\end{center}
\end{figure}

Fig.~\ref{mixN=Z}a shows the isospin mixing in $N=Z$ nuclei as a
function of the mass number $A$. Results obtained before and after
rediagonalization are shown by open and full dots, respectively. In
both variants of the calculations, the isospin mixing shows a gradual
increase as a function of $A$. It increases from a fraction of a
percent in $^{16}$O to about 4\%--5\%, depending on the variant of the
calculation. Note that the results obtained before
rediagonalization follow closely those obtained in
Ref.\cite{[Dob95a]}.

The isospin mixing obtained after removing the spurious
mean-field component through the Coulomb rediagonalization is systematically larger
than the one obtained within the HF method followed by the exact isospin
projection. This result confirms that the mean-field breaks the isospin
symmetry in such a way that it counterbalances the external symmetry breaking
mechanism caused by the Coulomb field. Nevertheless, as clearly seen in
Fig.~\ref{mixN=Z}b, the HF energy is astonishingly close to the total energy
obtained after the Coulomb rediagonalization.

\section{Summary}\label{summary}

In the present study, we presented a new theoretical tool that
allowed for isospin projection of Slater determinants and we
discussed first applications of the formalism to calculate the
isospin-mixing parameters $\alpha_C$ and total binding energies in
$N=Z$ nuclei. In particular, we discussed the basis-size dependence
of $\alpha_C$ and we showed that the basis truncation may introduce
about $\pm 4$\% uncertainty in $\alpha_C$.

We also discussed the role and magnitude of the spurious
isospin-symmetry-violating response of the self-consistent mean field
against the physical symmetry-breaking effects of the Coulomb field.
We showed that even in $N=Z$ nuclei, the self-consistent mean-field
may induce unphysical isospin mixing that reduces $\alpha_C$ by
as much as 30\% in $^{100}$Sn. This unphysical mechanism is due to
the very variational nature of the self-consistent mean-field scheme,
which introduces its own isospin-symmetry breaking field and partly
counterbalances the repulsive symmetry-breaking Coulomb field so as
to minimize the total binding energy. Nevertheless, our calculations
show that the HF binding energies follow extremely closely those
obtained by rediagonalizing the Hamiltonian within the set of isospin-projected
states.

This work was supported in part by the Polish Ministry of
Science under Contract No.~N~N202~328234 and
by the Academy of Finland and University of
Jyv\"askyl\"a within the FIDIPRO programme.


\begin{thebibliography}{10}

\bibitem{[Eng70]}
{C.A. Engelbrecht and R.H. Lemmer, Phys. Rev. Lett. {\bf 24}, 607 (1970)}.

\bibitem{[Cau80]}
{E. Caurier, A. Poves, and A. Zucker, Phys. Lett. {\bf 96B}, 11 (1980); Phys.
  Lett. {\bf 96B}, 15 (1980)}.

\bibitem{[Cau82]}
{E. Caurier and A. Poves, Nucl. Phys. {\bf A385}, 407 (1982)}.

\bibitem{[Aue83]}
{N. Auerbach, Phys. Rep. {\bf 98}, 273 (1983)}.

\bibitem{[Dob00d]}
{J. Dobaczewski and J. Dudek, Comput. Phys. Commun. {\bf 102}, 166 (1997); {\bf
  102}, 183 (1997); {\bf 131}, 164 (2000)}.

\bibitem{[Dob04fw]}
{J. Dobaczewski, J. Dudek, and P. Olbratowski, Comput. Phys. Comm. {\bf 158}
  (2004) 158; HFODD User's Guide nucl-th/0501008.}

\bibitem{[Zdu07]}
{H. Zdu{\'n}czuk, J. Dobaczewski, and W. Satu{\l}a, Int. J. Mod. Phys. E {\bf
  16}, 377 (2007)}.

\bibitem{[Zdu07a]}
{H. Zdu{\'n}czuk, W. Satu{\l}a, J. Dobaczewski, and M. Kosmulski, Phys. Rev.
  {\bf C76}, 044304 (2007)}.

\bibitem{[Har05]}
{J.C. Hardy and I.S. Towner, Phys. Rev. {\bf C71}, 055501 (2005); Phys. Rev.
  Lett. {\bf 94}, 092502 (2005); J.C. Hardy, hep-ph/0703165.}

\bibitem{[Tow08]}
{I.S. Towner and J.C. Hardy, Phys. Rev. {\bf C77}, 025501 (2008)}.

\bibitem{[Rin80]}
{P. Ring and P. Schuck, {\sl The Nuclear Many-Body Problem} (Springer-Verlag,
  Berlin, 1980)}.

\bibitem{[Var88]}
{D.A. Varshalovich, A.N. Moskalev, and V.K. Khersonskii, {\sl Quantum Theory of
  Angular Momentum} (World Scientific, Singapore 1988)}.

\bibitem{[Per04]}
{E. Perli\'nska, S.G. Rohozi\'nski, J. Dobaczewski, and W. Nazarewicz, Phys.
  Rev. C {\bf 69}, 014316 (2004)}.

\bibitem{[Gir83]}
{M. Girod and B. Grammaticos, Phys. Rev. {\bf C27}, 2317 (1983)}.

\bibitem{[Dob96]}
{J. Dobaczewski, W. Nazarewicz, T.R. Werner, J.-F. Berger, C.R. Chinn, and J.
  Decharg\'e, Phys. Rev. {\bf C53}, 2809 (1996)}.

\bibitem{[Dob05e]}
{J. Dobaczewski and J. Engel, Phys. Rev. Lett. {\bf 94}, 232502 (2005)}.

\bibitem{[Bei75fw]}
{M. Beiner, H. Flocard, N. Van Giai, and P. Quentin, Nucl. Phys. {\bf A238}, 29
  (1975).}

\bibitem{[Sat08a]}
{W. Satu{\l}a, J. Dobaczewski, W. Nazarewicz, and M. Rafalski, to be
  published}.

\bibitem{[Dob95a]}
{J. Dobaczewski and I. Hamamoto, Phys. Lett. {\bf 345B}, 181 (1995)}.

\end{thebibliography}

\end{document}